\documentclass[aps,pra,twocolumn,superscriptaddress,showpacs]{revtex4}

\usepackage{amsmath}
\usepackage{amsfonts}
\usepackage{amssymb}
\usepackage{multirow}
\usepackage{verbatim}
\usepackage{alltt}
\usepackage{moreverb}
\usepackage{graphicx,color,graphics}
\usepackage{hyperref}
\usepackage{setspace}
\usepackage{url}

\providecommand{\ignore}[1]{}

\begin{document}
\singlespacing

\title{Efficient quantification of experimental evidence against local realism}
\author{Yanbao Zhang}
 \affiliation{Department of Physics, University of Colorado Boulder, Boulder, 
 Colorado, 80309, USA}
 \affiliation{Applied and Computational Mathematics Division, National Institute 
 of Standards and Technology, Boulder, Colorado, 80305, USA}
\author{Scott Glancy}
 \affiliation{Applied and Computational Mathematics Division, National Institute 
 of Standards and Technology, Boulder, Colorado, 80305, USA}
 \author{Emanuel Knill}
 \affiliation{Applied and Computational Mathematics Division, National Institute 
 of Standards and Technology, Boulder, Colorado, 80305, USA}
 

\begin{abstract}
  Tests of local realism and their applications aim for very high
  confidence in their results even in the presence of potentially
  adversarial effects. For this purpose, one can measure a quantity
  that reflects the amount of violation of local realism and determine
  a bound on the probability, according to local realism, of obtaining
  a violation at least that observed.  In general, it is difficult to
  obtain sufficiently robust and small bounds. Here we describe an
  efficient protocol for computing such bounds from any set of Bell
  inequalities for any number of parties, measurement settings, or
  outcomes.  The protocol can be applied to tests of other properties
  (such as entanglement or dimensionality) that are witnessed by
  linear inequalities.
\end{abstract}

\maketitle

\section{Introduction}
\label{sect:introduction}
Theories designed according to ``local realism'' (LR) include a set of
hidden variables, which if known would predict all measurement
results; however, the values of the hidden variables cannot be
influenced by spacelike-separated events.  In 1964, Bell constructed an
inequality satisfied by all correlations accessible by LR and showed
that correlations between spacelike-separated measurements on two
quantum systems can violate this inequality~\cite{Bell}.  Since then,
many experimental tests showing Bell-inequality violations have been
performed (see Ref.~\cite{Genovese} for a review).  The importance of
such a test is twofold. First, it shows that local realistic (LR)
descriptions of bipartite quantum systems do not always exist. Second,
it supports quantum information tasks such as quantum key
distribution~\cite{Barrett2005, Masanes2009, Masanes2011} and
randomness generation~\cite{Pironio, Colbeck}.  Particularly in the
second case, a successful test is required to reject LR with very
high significance in the presence of adversarial effects.

To test a Bell inequality in an experiment, one needs to estimate the
probabilities of various outcomes from a finite number $N$ of
measurements. Due to uncertainties in the estimated probabilities, it
is conventional to present the violation of LR in terms of a typically
large number of experimental standard deviations of violation of a
Bell inequality.  While this provides information on the precision
with which a Bell-inequality violation is measured, it is not a valid
indicator of the strength of experimental evidence against
LR~\cite{Zhang2011}. For the latter, one needs to take into account
the possibility that $N$ data points generated by an LR model can
violate a Bell inequality due to statistical fluctuations in finite
samples. For a stronger conclusion, it is also necessary to account
for the possibility of adversarial variations of LR models in time~\cite{Barrett}.
Traditional data analysis methods are primarily intended for
characterizing precision and significance corresponding to a moderate
number of Gaussian standard deviations when the central limit theorem is
applicable (see App.~\ref{sect:statistics}).  In the other extreme, one 
can use large deviations theory to compute ``certificates'' for the 
violation of LR (see Refs.~\cite{Gill1, Gill2}). The conclusions of the 
former are weakened by distributional assumptions, while those of the latter 
can be overly pessimistic. Our approach is strictly valid in all relevant 
situations while maintaining the potential for asymptotic optimality and 
enabling device-independent comparisons of experiments.

Our approach can be viewed either as an implementation of the
framework of statistical $p$-values familiar from traditional
hypothesis testing, or as a refinement of large deviations principles
commonly used for deriving performance certificates in areas of
information science. To explain our approach, we consider the
former. In the context of interest, a $p$-value is the maximum
probability according to LR of obtaining a test statistic, such as a
Bell-inequality violation, at least as extreme as that observed.
Thus, a small $p$-value serves as a certificate against the
possibility of LR and as an implementation-independent measure of how
well the test performed.  Bounds on $p$-values support precise
statements on experimental evidence against LR and can be used as
security certificates in quantum key distribution and genuine-randomness 
generation.

There are two available protocols that compute upper bounds of
$p$-values. One is the martingale-based protocol~\cite{Gill1,Gill2},
but the bounds computed, such as the bound in Ref.~\cite{Pironio}, are
not tight~\cite{Zhang2011}. The other is the prediction-based-ratio
(PBR) protocol~\cite{Zhang2011}, which computes tighter bounds.
Specifically, the latter bounds are asymptotically tight with respect
to $N$, if the prepared quantum states and measurement settings do not
vary in time. While the PBR protocol is practical for many standard
configurations, it is inefficient with respect to the number of
parties per test, settings per party, and outcomes per setting.
Relevant examples to which the PBR protocol cannot be effectively applied 
include any configuration involving continuous variables.  Here, we 
propose a simplified PBR protocol to efficiently compute high-quality 
$p$-value bounds for all configurations.

The simplified PBR protocol has several advantages over other
protocols.  First, its $p$-value bounds are as good as and typically
better than those obtained by the martingale-based protocol.  Second,
it can take multiple Bell inequalities into consideration at once in a
rigorous way. Thus we can obtain high-quality $p$-value bounds even
when we cannot determine beforehand which inequality will work best.
Third, it can adapt to changes in the experimental results'
distribution. In particular, it is valid without making independence,
stability, or asymptotic assumptions, even in adversarial contexts.  
Fourth, this protocol can be applied to any test with linear witnesses, 
such as entanglement detection~\cite{Horodecki, Terhal}, without a full
analysis of the relevant probability space.

\section{Preliminaries}
\label{sect:preliminaries} 
An experimental test of LR involves a number 
of trials. At each trial, each of a number of spatially separated
parties performs a local measurement, where the setting is chosen
randomly from a fixed set.  Conventionally, at the end of the
experiment, a predetermined Bell inequality is tested using the
results from the trials. For example, if there are two parties and
each party has two measurements with outcomes $\pm 1$, the
Clauser-Horne-Shimony-Holt (CHSH) inequality~\cite{Clauser}
\begin{equation}
E(A_1B_1)+E(A_1B_2)+E(A_2B_1)-E(A_2B_2)\leq2\label{eq:CHSH}  
\end{equation}
can be tested, where $E(A_iB_j)$ with $i,j\in\{1,2\}$ is the
correlation between measurements $A_i$ and $B_j$.

To statistically quantify the evidence against LR, we express a Bell inequality 
as an upper bound on the expectation of a function of a trial result. 
That is, we write a Bell inequality in the form $\langle I(X)\rangle\leq B$, 
where $I$ is a real-valued function, called a Bell function,  and $X$ is the 
random variable from which a trial result $x$ is sampled. The result $x$ 
consists of the measurement-setting choices made by all parties and the 
outcomes of these measurements. To write a Bell inequality in the 
above form, the probability distribution of the joint measurement-settings
is assumed to be known and fixed before an experiment. There is no loss of 
generality in assuming this, as explained in Refs.~\cite{Gill1,Gill2,Zhang2011}. 
For example, for the CHSH inequality~\eqref{eq:CHSH}, a trial result $x$ consists of 
setting choices $i,j$ and outcomes $a_i, b_j$, and we can write 
$I_{\text{CHSH}}(x)=4(1-2\delta_{i,2}\delta_{j,2})a_ib_j$ and $B=2$, where 
we have assumed that the joint-setting distribution is uniform.

As explained in Sec.~\ref{sect:introduction}, we quantify the strength of experimental 
evidence against LR by means of a $p$-value. A $p$-value is associated 
with a test statistic $T$ that is a function of the sequence of trial
results. If $N$ is the total number of trials, the corresponding sequence of results
is denoted by $\mathbf{x}=(x_1,\ldots, x_N)$. As is conventional, we distinguish
between the sequence of results and the sequence of random variables $\mathbf{X}
=(X_1,\ldots, X_N)$ giving rise to these results. The exact $p$-value $p_N$ is 
defined as the maximum of the probabilities of the events
 $T(\mathbf{X}_\text{LR})\geq T(\mathbf{x})$ over all random-variable sequences 
$\mathbf{X}_{\text{LR}}$ distributed according to LR models. That is,
\begin{equation}
p_N= \max_{\text{LR}} \text{Prob}_\text{LR}(T(\mathbf{X}_{\text{LR}})\geq T(\mathbf{x})).  \label{eq:p-value}
\end{equation}
Due to the difficulty of determining worst-case tail probabilities of typical 
test statistics, we can usually determine only upper bounds of exact $p$-values. 
Thus, for the remainder of the paper, the term ``$p$-value'' refers to any valid 
upper-bound on the exact $p$-value. For the protocols discussed below, if the 
trial results are independent and identically distributed according to a 
distribution that violates LR, the $p$-values computed decrease to $0$ 
exponentially as $N\to\infty$. We can therefore compare different protocols'
performances in a test of LR according to the confidence-gain rate defined by  
\begin{equation}
G=-\lim_{N\to\infty} \frac{\log_2 p^{\text{(prot)}}_{N}}{N},  \label{eq:gain_rate} 
\end{equation}
where $p^{\text{(prot)}}_{N}$ is the $p$-value computed by a protocol. Higher gain 
rates imply better protocol performance. Each protocol discussed below works even 
under memory effects~\cite{Barrett}, that is, even when the prepared 
quantum state, measurement settings, and relevant LR models vary arbitrarily 
with time. 

\section{PBR protocols}
\label{sect:pbr_protocols}
The test statistic used by a PBR protocol is based on 
non-negative functions $R_n$ to be applied to the $n$'th trial result $x_n$ and
satisfying $\langle R_n(X)\rangle\leq 1$ for $X$ distributed according to any LR 
model. Thus, each $R_n$ is a non-negative Bell function for the Bell inequality 
$\langle R_n(X)\rangle\leq 1$. The function $R_n$ is constructed \emph{before} observing 
the $n$'th trial result $x_n$. Its construction can use information from previous 
trials and typically requires predicting the distribution of $X_n$. Thus, $R_n$ is 
referred to as a PBR. A PBR protocol computes a test statistic according to 
$T(\mathbf{x})=\prod_{n=1}^N R_{n}(x_n)$.  To obtain a $p$-value for $T(\mathbf{x})$, 
it suffices to observe that by construction $T$ is non-negative and 
$\langle T(\mathbf{X}_\text{LR})\rangle\leq 1$, so that by Markov's inequality we 
can compute a $p$-value according to
\begin{equation}
p_N^{(\text{PBR})}=\min(1/T(\mathbf{x}),1). \label{eq:pbr_p-value}
\end{equation} 
See Ref.~\cite{Zhang2011} for further details. Different PBR protocols are 
characterized by how they choose the PBR $R_n$ for each $n$. 

\subsection{Full PBR protocol}
\label{sect:full_pbr_protocol}
For the full PBR protocol~\cite{Zhang2011}, $R_n$ is 
chosen so as to optimize the expected confidence-gain rate given previous trial results.
For this optimization, the protocol assumes that $X_n$'s distribution is the same as 
that from which the trial results $x_1,\ldots,x_{n-1}$ were sampled, and that these 
samples are independent. Whether or not these assumptions actually hold affects only 
the quality of the $p$-value computed, but not its validity. Given these assumptions, 
$R_n$ is computed in two steps. The first is to make an estimate of the experimental 
probability distribution $q(x)\equiv\text{Prob}_{\text{QM}}(X_n=x)$, and the second 
is to determine the probability distribution $p(x)\equiv\text{Prob}_{\text{LR}}(X_\text{LR}=x)$ 
according to the LR model that minimizes the Kullback-Leibler (KL) divergence from 
the estimate $q$~\cite{Kullback}. The protocol then sets the next PBR to 
$R_n(x_n)=q(x_n)/p(x_n)$. Details about this protocol and the proof that this 
$R_n$ satisfies the conditions on a PBR are in Ref.~\cite{Zhang2011}.

\subsection{Simplified PBR protocol}
\label{sect:simplified_pbr_protocol}
The simplified PBR protocol chooses the PBRs 
from the convex combinations of Bell functions that are derived from a 
given set of Bell inequalities. To ensure that a convex combination is a PBR, the 
Bell functions first need to be standardized so that they are non-negative and have 
expectations at most $1$ for any LR model. Any Bell function that is lower-bounded 
has such a standardized form. In particular, if $\langle I(X)\rangle\leq B$
is a Bell inequality and $I(x)\geq b$ for all $x$, then $r(x)=(I(x)-b)/(B-b)$ 
is standardized. Note that, as a constraint on the distribution of $X$, 
$\langle r(X)\rangle\leq 1$ is equivalent to $\langle I(X)\rangle\leq B$. Given 
Bell inequalities $\langle I^{(m)}(X)\rangle\leq B^{(m)}$, where $I^{(m)}$ is 
lower-bounded and $m=1,2,\ldots,M$, we can construct the corresponding
standardized Bell functions $r^{(m)}$. We define $\mathbf{r}=(r^{(1)},\ldots,r^{(M)})$.  
The simplified PBR protocol chooses the PBR $R_n$ from among the convex combinations
\begin{equation}
\boldsymbol{\omega}\cdot \mathbf{r}=\sum_m \omega_m r^{(m)}, \label{eq:simplified_pbr}
\end{equation}
where $\omega_m\geq 0$ and $\sum_m \omega_m=1$. Our implementation
always includes the trivial Bell function $r^{(1)}=1$. This ensures
that the set of convex combinations is at least one-dimensional and
that the confidence-gain rate is at least as high as that achieved
by the martingale-based protocol.

Like the full PBR protocol, the simplified PBR protocol aims to
optimize the expected confidence-gain rate given previous trial
results, under the assumption that the distribution of $X_n$ is the
same as the empirical-frequency distribution of the previous trial
results.  Whether or not this assumption holds does not affect the
validity of the $p$-value computed.  The confidence gain of the $n$'th
trial may be defined as $\log_2 R_n(x_n)$.
Its expected value given that $X_n$ is distributed according to $q$ is 
\begin{equation}
\sum_{x_n}q(x_n)\log_2 R_n(x_n).
\end{equation}
Before the $n$'th trial, the protocol attempts to maximize this
expected confidence gain. Since $q$ is not known, it is empirically
estimated based on the results observed at the previous $(n-1)$ trials.  
Expanding $R_n$ according to Eq.~\eqref{eq:simplified_pbr} yields the 
following estimate of the expected confidence gain of the $n$'th trial:
\begin{equation}
G_n(\boldsymbol{\omega})=\frac{1}{n-1}\sum_{k=1}^{n-1}\log_2(\boldsymbol{\omega} \cdot \mathbf{r}(x_k)).
\label{eq:optimal_choice}
\end{equation}
The protocol thus determines $R_n$ by maximizing
$G_n(\boldsymbol{\omega})$ over $\boldsymbol{\omega}$, that is, $R_n=\mathbf{r}\cdot\mathrm{argmax}_{\boldsymbol{\omega}}G_n(\boldsymbol{\omega})$.
Note that, unlike the full PBR protocol, the simplified PBR protocol
does not require explicitly optimizing over all LR models.  Computing
$\mathrm{argmax}_{\boldsymbol{\omega}}G_n(\boldsymbol{\omega})$ requires
optimizing a convex objective function over an $M$-dimensional
convex space. In our implementation, we apply the expectation-maximization 
(EM) algorithm~\cite{Cover} to solve this problem.

Before an experiment, it is important to choose a relevant (and preferably 
small) set of Bell functions based on the quantum state prepared and other 
experimental parameters.  Expanding the set of Bell functions can increase 
the confidence-gain rate at the cost of increased computation.  (The optimal 
gain rate can be achieved if the best Bell function according to the full PBR
protocol is included in the chosen set~\cite{Zhang2011}.) Below we show that 
it helps to include more than just the obvious Bell functions.

The performance of the simplified PBR protocol can be compared with that of the 
martingale-based protocol~\cite{Gill1, Gill2}, the only valid non-PBR protocol 
considered so far. The martingale-based protocol uses a Bell 
inequality $\langle I(X)\rangle\leq B$ with a bounded Bell function, chosen before 
an experiment. After the experiment, the mean of the Bell function is estimated 
as $\hat{I}=\frac{1}{N}\sum_{n=1}^N I(x_n)$. To obtain a $p$-value, the protocol 
uses $\hat{I}$ as the test statistic. If $\hat{I}\geq B$, the bounds on the Bell 
function imply that a $p$-value can be computed according to 
\begin{equation}
p_N^{(\text{mart})}=\left[\left(\frac{a-B}{a-\hat{I}}\right)^{\frac{a-\hat{I}}{a-b}}\left(\frac{B-b}{\hat{I}-b}\right)^{\frac{\hat{I}-b}{a-b}}\right]^N, \label{eq:mart_p-value}
\end{equation}
where $a=\sup_xI(x)$ and $b=\inf_xI(x)$. This $p$-value expression is based on a 
version of Hoeffding's bound in Refs.~\cite{Hoeffding, mcdiarmid:qc1989a}, which 
improves the ones given in Refs.~\cite{Pironio, Zhang2011, Gill2}. The derivation of Eq.~\eqref{eq:mart_p-value} is explained in App.~\ref{sect:derivation}. In 
App.~\ref{sect:proof}, we show that the 
simplified PBR protocol using the same Bell inequality, together with the default 
trivial Bell function $r=1$, achieves a gain rate at least as high as the gain 
rate achieved by the martingale-based protocol. These two gain rates are equal 
to each other if and only if the experimental range of the function $I$ is 
contained in the set $\{a,b\}$.  

\begin{figure}[tb!]
   \includegraphics[scale=0.56, viewport=3.5cm 10cm 18cm 19.8cm]{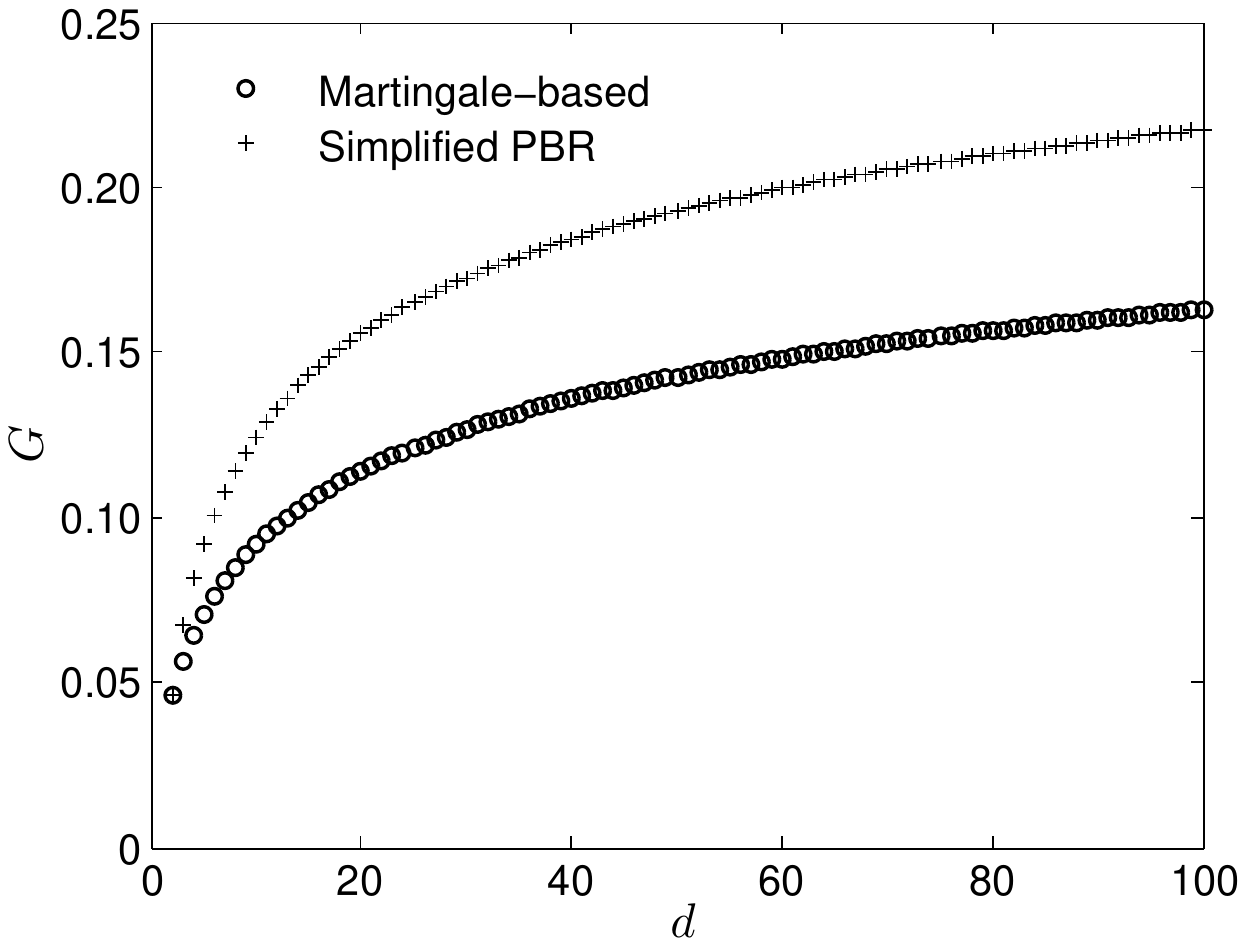}
   \small{
   \caption{Confidence-gain rates in the test of the CGLMP inequality 
   $\langle I_d(X)\rangle\leq 2$. Here, we use the quantum state and measurement 
   settings of Ref.~\cite{Chen}, Eqs.~(15) and~(9), respectively. 
   }\label{fig:CGLMP}}
\end{figure}

\subsection{Computational resource costs}
\label{sect:resource_costs}
Of the available protocols for computing $p$-values, 
the martingale-based one is the least resource-intensive and simplest to apply.  It requires 
computing only an estimate of the mean of the Bell function, which involves a sum of $N$ terms.

To assess the computational resource cost of the simplified PBR
protocol, we assume that optimizations are performed by algorithms whose
resource costs are determined primarily by the complexity $C$ of
evaluating the objective function and the dimension $D$ of the convex
search space. Given this and other assumptions detailed in App.~\ref{sect:efficiency} 
and from Eq.~\eqref{eq:optimal_choice}, the complexity of evaluating the
objective function in the optimization required for constructing the
PBR $R_n$ is $C=O(nM)$, which is the product of the number of terms in
the sum and the number of Bell-function evaluations underneath the
logarithm.  The dimension of the search space is $D=O(M)$, the size of
$\boldsymbol{\omega}$ in Eq.~\eqref{eq:optimal_choice}.  Consequently,
unlike the full PBR protocol (whose computational resource cost is
detailed in App.~\ref{sect:efficiency}), the numbers of parties, settings, 
and outcomes are not limiting factors. In this sense, the simplified PBR 
protocol is efficient for any experimental configuration.

\section{Protocol comparison}
\label{sect:comparison} 
We begin by comparing the confidence-gain rates 
achieved by different protocols for experimental configurations designed to 
violate the Collins-Gisin-Linden-Massar-Popescu (CGLMP) inequality~\cite{Collins}.
To test the CGLMP inequality, there are two parties, and each of them
performs one of two possible measurements with $d$ outcomes at each
trial.  This is an example where the full PBR protocol is impractical
for large $d$. For this example and the one below, we assume that at
each trial each party's measurement setting is chosen uniformly
randomly. The CGLMP inequality can be written as $\langle
I_d(X)\rangle\leq 2$, where the function $I_d$ takes $d$ different
values.  The gain rates $G_\text{mart}$ and $G_\text{sPBR}$, achieved
by the martingale-based and simplified PBR protocols, are shown in
Fig.~\ref{fig:CGLMP}.  Here the simplified PBR protocol uses only the
CGLMP inequality.  This figure illustrates that $G_{\text{sPBR}}$ is
higher than $G_{\text{mart}}$ when $d>2$.  

The optimal gain rate $S_q$ is achieved by the full PBR protocol and
can be computed as the minimum KL divergence from the experimental
probability distribution to any LR model~\cite{Bahadur}. For the
results of Fig.~\ref{fig:CGLMP}, we find that the gain rates
$G_{\text{sPBR}}$ are numerically indistinguishable from $S_q$ when
$d\leq13$. For the case $d>13$, it is difficult to compute $S_q$ due
to the large dimension of the probability space over all possible LR
models. For the tests studied in Fig.~\ref{fig:CGLMP}, we conjecture
that $G_{\text{sPBR}}=S_q$.  In general, we cannot guarantee that
$G_{\text{sPBR}}$ is optimal.

\begin{figure}[tb!]
   \includegraphics[scale=0.49, viewport=3cm 9cm 18.5cm 20.5cm]{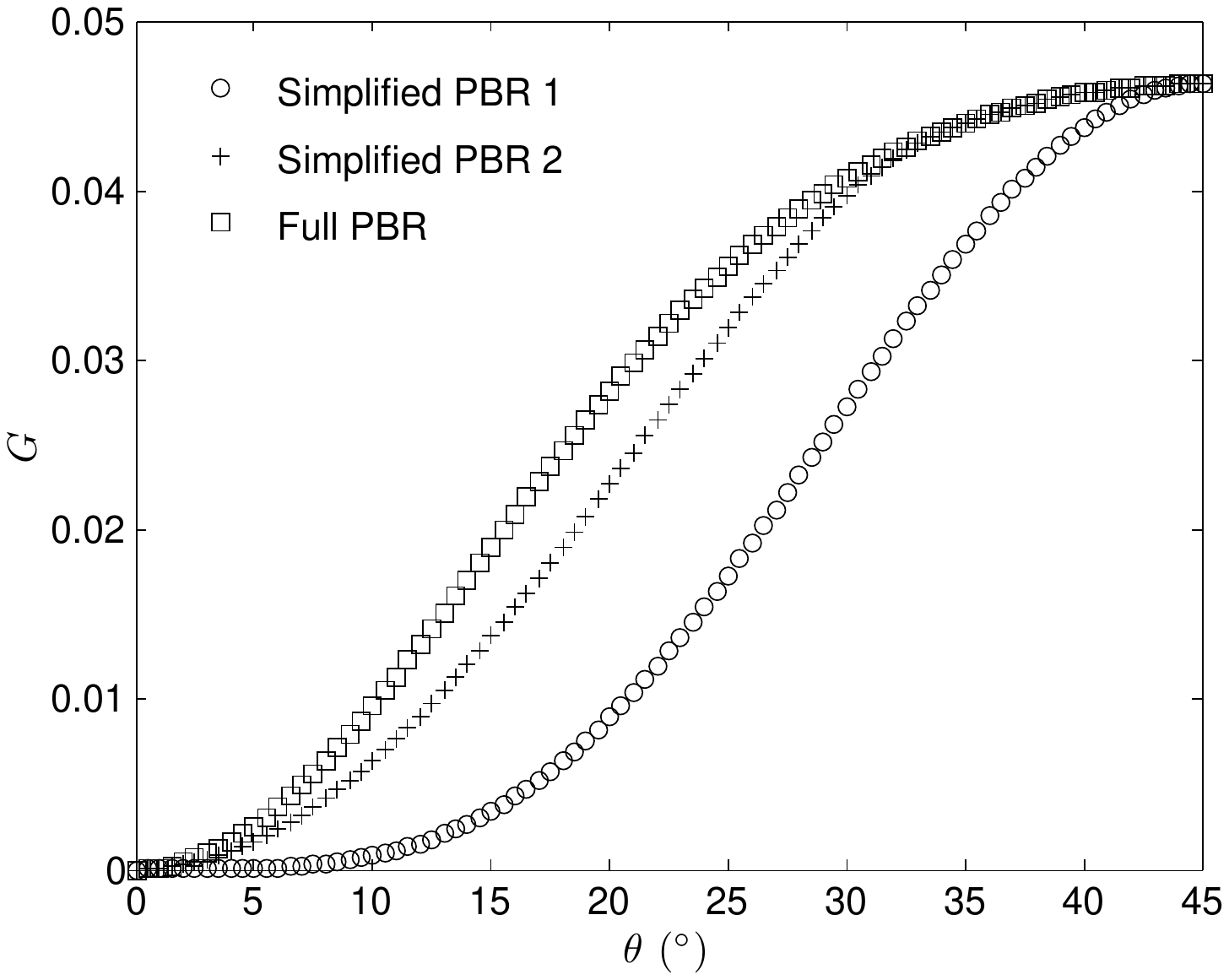}
   \small{
   \caption{Confidence-gain rates in the test of LR with an unbalanced Bell 
   state $|\psi(\theta)\rangle$. The measurement settings are chosen to 
   maximize the violation of the CHSH inequality~\eqref{eq:CHSH} given 
   the state $|\psi(\theta)\rangle$. The gain rates achieved by the simplified 
   PBR protocol using the CHSH inequality are shown as circles ($\circ$), while 
   the gain rates by the same protocol using the CHSH inequality together with 
   no-signaling conditions are shown as crosses ($+$).}
   \label{fig:CHSH}}
\end{figure}

Next, we compare the performance of the simplified PBR protocol when
using different numbers of Bell inequalities. The experimental
configuration considered is for a test of the CHSH
inequality using an unbalanced Bell state 
$|\psi(\theta)\rangle = \cos(\theta)|00\rangle +
\sin(\theta)|11\rangle$. For comparison, we consider the simplified 
PBR protocol with the CHSH inequality~\eqref{eq:CHSH} alone or in conjunction 
with additional, seemingly trivial Bell inequalities such as those derived 
from no-signaling conditions. With Bell functions corresponding to
no-signaling conditions, the gain rates are improved, as shown in
Fig.~\ref{fig:CHSH}.

In App.~\ref{sect:finite behaviour}, we show how the $p$-values computed by  
different protocols behave in a simulated experiment as functions of the number 
of trials.

\section{Extensions}
\label{sect:extensions} 
To compute a $p$-value, the simplified PBR 
protocol uses a set of linear inequalities that are satisfied by the
predictions of a null hypothesis before each trial in an
experiment. Besides tests of LR, there are many other types of tests
based on linear witnesses, such as tests for
entanglement~\cite{Horodecki, Terhal} and system dimensionality above
a given bound~\cite{Brunner, Gallego}. In any test based on linear
witnesses, such a witness can be expressed as $\langle W(X)\rangle\leq
B$, where $W$ is a real-valued function and $X$ is the random variable
from which a trial result $x$ is sampled. The result $x$ consists of
all choices made at each trial, such as choices of states and
measurement settings, and the outcomes observed under these choices.
Here, we assume that the choices are made randomly according to a
known probability distribution at each trial, so that a witness
$\langle W(X)\rangle\leq B$ is satisfied before each trial assuming
the null hypothesis. As for Bell functions, if a witness function $W$ is
lower-bounded it can be standardized. The simplified PBR protocol can
then be applied with any set of standardized witnesses, as we did in a
test of LR.

\begin{acknowledgments}
We thank Kevin Coakley for helpful comments and discussions.
This paper is a contribution of the National Institute of Standards 
and Technology and is not subject to U.S. copyright.
\end{acknowledgments}

\appendix*
\section{}
\label{sect:appendix}

\subsection{Computational resource comparison}
\label{sect:efficiency}
In this appendix, we compare the computational resources required by the simplified 
and full PBR protocols in an experimental test of LR. 

We consider an experimental configuration involving $l$ parties where each party has 
$s$ measurement settings and each local measurement has $d$ outcomes. (The comparison 
below is readily extended to more general configurations.) We suppose that the 
joint-setting distribution is uniform. Then, the number of possible results 
at a trial is $K=(ds)^l$. 
Since a deterministic LR model specifies the exact outcome 
for each local measurement of each party at a trial, there are $H=d^{ls}$ many such 
models. A general LR model is a convex combination of deterministic LR models, so the 
number of free parameters characterizing a general LR model is $(H-1)$.

Let the total number of trials in an experimental test of LR be $N$. We assume 
that each PBR protocol sets the initial value of the PBR to 
$R_1=1$ and updates the PBR $R_n$ before each trial $n$ ($n>1$). (In practice this 
is unnecessary; see the appendix of Ref.~\cite{Zhang2011}.) For updating the PBR, 
each PBR protocol needs to optimize a convex objective function over a convex space. 
The complexity of this optimization problem can be described in terms of variables 
that are functions of the parameters $n$, $l$, $s$, and $d$ characterizing the input 
data size. (Note that the stored size of the first $n$ trial results is $O(n \log(K)) 
= O(n l (\log(d)+\log(s)))$.) We need to quantify the resource cost of implementing 
each protocol in terms of these parameters.

The complexity of the optimization problem solved before each trial can be parametrized 
by the complexity of the convex search space, the complexity of evaluating the objective 
function, and the precision needed for computing a high-quality $p$-value for rejecting 
LR. We assume that the simplified and full PBR protocols use generic iterative optimization 
algorithms whose implementation complexities as functions of these parameters are 
asymptotically the same. We also assume that the complexity of the convex search space is 
dominated by its dimension. In particular, we do not account for the complexity of enforcing 
convex constraints. This is motivated by the observation that there is no additional overhead 
for enforcing convex constraints in the EM algorithm~\cite{Cover} 
used in our implementation. For quantifying the complexity of evaluating the objective function, 
we assume that the Bell functions used can be evaluated in constant time given any trial result. 
This assumption is realistic for many Bell functions, as their values are determined by concise
formulas derived from theory.  Alternatively, these functions can be preprocessed as a table 
stored in random-access memory; we do not include preprocessing time in our analysis. 
Also, we assume that determining whether or not an arbitrary trial result $x$ happens 
according to a deterministic LR model takes constant time. (Strictly speaking, the 
time taken for such a determination process is proportional to the number of parties $l$.) 
The precision needed affects the number of iterations required by an algorithm to find a 
numerical solution. It affects only the quality of the $p$-value computed by a protocol, 
but not its validity. (For the EM algorithm used, see Theorem $4$ of Ref.~\cite{Cover} and 
the appendix of Ref.~\cite{Zhang2011} for the effects of the precision parameters in the 
simplified and full PBR protocols, respectively.) We assume that the precision parameters 
in both protocols are set to be the same, and we do not account for the number of iterations 
required to achieve the specified precision.  Therefore, for the purpose of comparing the 
computational resources required by the simplified and full PBR protocols, we focus on 
comparing the dimensions $D$ of the convex search spaces and the complexities $C$ 
of evaluating the objective functions in the optimization problems solved by 
the two protocols before each trial.

We first consider the simplified PBR protocol. Given a set of $M$ Bell inequalities, 
this protocol sets $R_n=\boldsymbol{\omega}_n\cdot \mathbf{r}$, where the size of 
$\boldsymbol{\omega}_n$ is $M$, $\mathbf{r}$ is defined before Eq.~\eqref{eq:simplified_pbr},  
and $\boldsymbol{\omega}_n$ is chosen to maximize the estimated confidence-gain rate
\begin{equation}
\frac{1}{n-1}\sum_{k=1}^{n-1} \log_2 (\boldsymbol{\omega}\cdot \mathbf{r}(x_k))=\sum_{x: f_n(x)\neq 0} f_n(x) \log_2(\boldsymbol{\omega}\cdot \mathbf{r}(x)), \label{eq:optimal_spbr}
\end{equation}
where $f_n(x)$ is the empirical frequency of $x$ before the $n$'th trial. Note 
that, in the right-hand side of Eq.~\eqref{eq:optimal_spbr}, the sum is taken over 
only the results $x$ already observed in the previous trials.

For the maximization of Eq.~\eqref{eq:optimal_spbr}, the dimension of the convex 
search space is $M$. The evaluation of the objective function can use the left-hand
or right-hand side of Eq.~\eqref{eq:optimal_spbr}, whichever has fewer terms. Thus 
it involves a sum of at most $\min (n-1,K)$ terms where each term requires computing 
a convex combination of $M$ Bell-function values. Hence, for updating the PBR $R_n$ 
before the $n$'th trial, the complexity of evaluating the objective function is 
$C_{\text{sPBR}}=O(\min(nM,KM))=O(\min(nM,(ds)^lM))$, and the dimension of the 
search space is $D_{\text{sPBR}}=O(M)$. Therefore, if any of the configuration 
parameters $l$, $s$, or $d$ is large, $C_{\text{sPBR}}$ and $D_{\text{sPBR}}$ are 
independent of these parameters, and so the simplified PBR protocol can be applied 
efficiently.

The full PBR protocol~\cite{Zhang2011} computes $R_n$ in two steps. First, the protocol
estimates the probability $q(x)$ of the result $x$ to be observed at the next trial.  
This estimate can be obtained in different ways. The simplest is to let $q(x)$ be the 
empirical frequency $f_n(x)$ of $x$ over the previous $(n-1)$ trials. However, 
one can consider additional constraints such as the known joint-setting distribution 
and no-signaling conditions. Thus, in Ref.~\cite{Zhang2011} we suggested
maximizing the log-likelihood function $L(q')\propto \sum_x f_n(x) \log_2(q'(x))$, 
subject to these constraints, and we observed that this can improve the quality of 
the $p$-value computed. Since this maximization is not a resource bottleneck, we do 
not consider its complexity in the comparison. Second, we find the LR model $p$ 
closest to the estimated distribution $q$ by minimizing the KL 
divergence~\cite{Kullback} from $q$ to an LR model $p_\textrm{LR}$
\begin{equation}
D_{\text{KL}}(q|p_\textrm{LR})=\sum_x q(x)\log_2\frac{q(x)}{p_\textrm{LR}(x)}. \label{eq:kl_divergence}
\end{equation}
The full PBR protocol then sets $R_n(x_n)=q(x_n)/p(x_n)$. 

For the minimization of Eq.~\eqref{eq:kl_divergence}, the dimension of the convex 
search space is $H$. The evaluation of the objective function involves a sum of $K$ 
terms where each term requires computing $p_\textrm{LR}(x)$ according to a convex 
combination of $H$ deterministic LR models.  Hence, for updating the PBR
$R_n$ before the $n$'th trial, the complexity of evaluating the objective
function is $C_{\text{fPBR}}=O(KH)=O(d^{l(s+1)}s^l)$, and the dimension of the
search space is $D_{\text{fPBR}}=O(H)=O(d^{ls})$. While $C_{\text{fPBR}}$ 
and $D_{\text{fPBR}}$ are polynomial in $d$, they are exponential in each of $l$ 
and $s$. Therefore, the full PBR protocol is not efficient with respect to these
configuration parameters. 

Before applying the simplified PBR protocol, one chooses a relevant and preferably 
small set of Bell inequalities. In many cases of interest, $l$, $s$, or $d$ is 
large, and so is $H=d^{sl}$. For example, in field-quadrature measurements, $d$ is 
fundamentally infinite. Hence, $M$, the number of Bell inequalities used in the 
simplified PBR protocol, is in general much smaller than $H$, the number of 
deterministic LR models considered in the full PBR protocol. The complexities 
show that for such cases, the simplified PBR protocol is substantially less 
resource-intensive than the full PBR protocol.

\subsection{The martingale-based protocol's $p$-value} 
\label{sect:derivation}
Consider a Bell inequality $\langle I(X)\rangle\leq B$ with a Bell function 
$I$ whose range is included in the interval $[b,a]$, where $b\leq a$. 
An experimental test yields an estimate $\hat{I}=\frac{1}{N}\sum_{n=1}^{N} I(x_n)$ 
of the mean of $I$, where $x_1,\ldots,x_N$ are the trial results. 

Suppose that the $n$'th trial result $x_n$ is distributed according to a random 
variable $X_{\text{LR},n}$ satisfying LR. In this case, the random variable from 
which $\hat{I}$ is sampled is $I_\text{LR}=\frac{1}{N}\sum_{n=1}^{N} I(X_{\text{LR},n})$. 
The sequence $M_n=\sum_{k=1}^n(I(X_{\text{LR},k})-B)$, $n=1,\ldots,N$, is a super-martingale, 
as shown in Refs.~\cite{Gill1,Gill2}. Thus, for $t\geq0$, the probability
\begin{align}
\text{Prob}_\text{LR}&\left( M_N \geq Nt\right)\leq \notag \\ 
& \left[\left(\frac{a-B}{a-B-t}\right)^{\frac{a-B-t}{a-b}}\left(\frac{B-b}{B+t-b}\right)^{\frac{B+t-b}{a-b}}\right]^N.
\label{eq:mart_bound}
\end{align}
The inequality~\eqref{eq:mart_bound} follows from Theorem $6.1$ of 
Ref.~\cite{mcdiarmid:qc1989a}.  Since $\text{Prob}_\text{LR}(I_{\text{LR}} 
\geq \hat{I})=\text{Prob}_\text{LR}(M_N \geq N(\hat{I}-B))$, from the above 
inequality~\eqref{eq:mart_bound} we get the $p$-value of Eq.~\eqref{eq:mart_p-value}.  

Note that, although Theorem $6.1$ of Ref.~\cite{mcdiarmid:qc1989a} is 
stated for a martingale, the same result and its proof also apply to a 
super-martingale. The same bound is also derived in Theorem $1$ 
of Ref.~\cite{Hoeffding} for a sum of independent random variables. 
From Refs.~\cite{mcdiarmid:qc1989a, Hoeffding}, we can see that 
the bound in Eq.~\eqref{eq:mart_bound} is tighter than bounds of 
$\text{Prob}_\text{LR}\left( M_N \geq Nt\right)$ used in previous 
works~\cite{Gill2, Pironio, Zhang2011} and derived from Azuma's 
inequality~\cite{Azuma, mcdiarmid:qc1989a}.

\subsection{Proof of $G_\text{mart}\leq G_\text{sPBR}$} 
\label{sect:proof}
We suppose that the martingale-based protocol uses a Bell inequality 
$\langle I(X)\rangle\leq B$ with a bounded Bell function such that
$b\leq I(x)\leq a$ for all $x$. Also, we suppose that the simplified 
PBR protocol uses the standardized form of this Bell inequality together
with the trivial Bell function $r=1$. 

Let the experimental probability of observing the result $x$ be $q(x)$. 
The experimental mean of $I$ is $I_q=\int q(x)I(x)\text{d}x$. If 
$I_q\geq B$, then from Eqs.~\eqref{eq:gain_rate} and~\eqref{eq:mart_p-value} 
we get the gain rate
\begin{align}
& G_\text{mart}=\frac{a-I_q}{a-b}\log_2 \frac{a-I_q}{a-B}+\frac{I_q-b}{a-b}\log_2 \frac{I_q-b}{B-b} \notag \\
&=\int \left(\frac{a-I(x)}{a-b} \log_2 \frac{a-I_q}{a-B} + \frac{I(x)-b}{a-b} \log_2 \frac{I_q-b}{B-b} \right) \notag \\
&\qquad q(x)\text{d}x . \label{eq:mart_gain1}
\end{align}
Here, we use the fact that the experimental estimate $\hat{I}$ approaches $I_q$ as 
$N\rightarrow\infty$. By the concavity of $\log_2(x)$ and some algebra, we get that 
the gain rate $G_\text{mart}$ satisfies the inequality
\begin{align}
G_\text{mart} &\leq \int  \log_2\left(\frac{a-I(x)}{a-b}  \frac{a-I_q}{a-B} + \frac{I(x)-b}{a-b} \frac{I_q-b}{B-b} \right) \notag \\
&\qquad q(x)\text{d}x \notag \\
&=\int \log_2\left(\omega_0 \frac{I(x)-b}{B-b} +1-\omega_0 \right) q(x)\text{d}x, \label{eq:mart_gain2}
\end{align}
where $0\leq\omega_0=\frac{I_q-B}{a-B}\leq1$.

From Eqs.~\eqref{eq:gain_rate} and~\eqref{eq:pbr_p-value} and according to the design of the PBRs 
by the simplified PBR protocol (as explained in Sec.~\ref{sect:simplified_pbr_protocol}),  
the gain rate achieved by this protocol is
\begin{equation}
G_\text{sPBR}=\max_{0\leq\omega\leq1}\int  \log_2\left(\omega \frac{I(x)-b}{B-b} +1-\omega \right)
q(x) \text{d}x. \label{eq:spbr_gain}
\end{equation} 
Here, we use the fact that the empirical frequency $f_N(x)$ approaches the experimental 
probability $q(x)$ as $N\rightarrow\infty$. The inequality $G_\text{mart}\leq G_\text{sPBR}$ 
follows from comparing Eq.~\eqref{eq:mart_gain2} with Eq.~\eqref{eq:spbr_gain}.

By considering the condition for the equality in Eq.~\eqref{eq:mart_gain2}, 
we can show that $G_\text{mart}=G_\text{sPBR}$ if and only if $q(x)=0$ 
whenever $b<I(x)<a$. For this it suffices to note that $\log_2(x)$ is strictly 
concave, so the equality in Eq.~\eqref{eq:mart_gain2} holds if and only if
$I(x)=a$ or $b$ whenever $q(x)\neq0$.

\subsection{Behavior of the protocols for finite data} 
\label{sect:finite behaviour}
Here we consider the behavior of each protocol given a finite amount of experimental 
data. We simulate the test of the CGLMP inequality 
$\langle I_3(X)\rangle \leq2$~\cite{Collins} with the quantum state  
and measurement settings of Ref.~\cite{Chen}, Eqs.~(15) and~(9) (with $d=3$),  
respectively. We assume that at each trial each party's measurement setting is 
chosen uniformly randomly. The protocols' gain rates are $G_{\text{mart}}=0.0565$ 
and $G_{\text{sPBR}}=0.0675$, while the optimal gain rate $S_q$ achieved by the full 
PBR protocol is numerically indistinguishable from $G_{\text{sPBR}}$. For computing 
$G_{\text{sPBR}}$, the simplified PBR protocol uses the standardized CGLMP inequality 
and the trivial Bell function $r=1$. 

\begin{figure}[htb!]
   \includegraphics[scale=0.56, viewport=3.5cm 8.5cm 18.5cm 19.5cm]{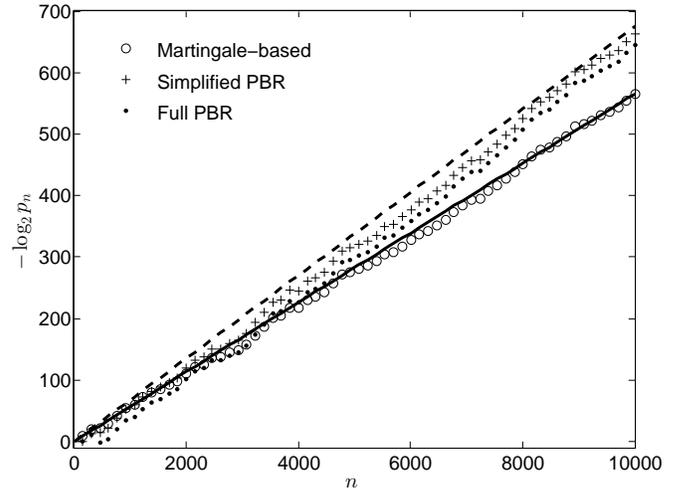}  
   \small{
   \caption{An example of running log-$p$-values as functions of the number of
     trials $n$ in a test of the CGLMP inequality. The dashed and solid lines 
     are the asymptotic lines for log-$p$-values based on gain rates achieved by the 
     (full or simplified) PBR protocol and the martingale-based protocol, respectively. 
     Repetitions of this Monte Carlo simulation show similar behavior.} 
   \label{fig:simulation}}
\end{figure} 

The results from $10,000$ successive trials are recorded. 
Fig.~\ref{fig:simulation} shows the (negative) log-$p$-values computed for 
the first $n$ results from a simulated sequence of trials as functions of $n$. 
The asymptotic lines for log-$p$-values, given by the products of $n$ and  
the respective gain rates achieved by different protocols, are also shown in 
Fig.~\ref{fig:simulation}. 

In our discussion so far, we have assumed that each PBR protocol updates the 
PBR before each trial. In practice, the PBR is updated only for a block of 
trial results at a time. Specifically, for the simulation shown in 
Fig.~\ref{fig:simulation}, we update the PBRs and log-$p$-values only after 
every block including $154$ successive trials. (See our previous 
work~\cite{Zhang2011} for a discussion of the block-size choice and related 
issues.) This block-size choice limits PBR computations to when enough new 
information has been obtained, thereby reducing the resource cost. It also 
mitigates the offset of the computed log-$p$-values from the asymptotic line. 
This offset is due to an initial transient where the relevant features 
of the experimental distribution are being learned. The learning offset can be 
removed if, before an experiment, we have a good estimate of the experimental 
results' distribution. Such an estimate could be based on (quantum or otherwise) 
theory or previous experiments. 

The PBR protocols provide better results than the martingale-based protocol. 
However, the PBR log-$p$-values show learning offsets from the asymptotic line. 
Our results show that the simplified PBR log-$p$-values have a smaller learning 
offset than the full PBR log-$p$-values in each of $30$ independent simulations 
performed. The reason is that the simplified PBR protocol needs to infer a much 
smaller number of parameters for constructing the PBRs. 
  
In the above example, the simplified PBR protocol uses only two Bell
functions.  Given a prescient choice of Bell functions, this is
sufficient for computing asymptotically optimal $p$-values. But in 
general, more Bell functions are needed for computing a
  high-quality $p$-value. However, this involves inferring more
parameters and thus requires more trials before a good inference
can be obtained. As a result, the learning offset is expected to
  increase when using more Bell functions. One way to mitigate this
problem may be to increase the number of Bell functions used over
time, adding new Bell functions only when there are enough trials for
reliable inference of the additional parameters.

\subsection{Statistical Issues}
\label{sect:statistics}

There are standard statistical approaches to testing composite null
hypotheses such as the set of LR models. These approaches generally assume that every
model in the set of null hypotheses can be associated with a definite
probability distribution for the trial results and that one has a
parametrization of alternative hypotheses. They then require that one
has access to the results from a large number of independent and
identical trials, on which one computes a test statistic.  A good
example of a standard approach is the likelihood-ratio test. (See
\cite{Robin} for an entanglement test based on a likelihood ratio.)
In addition, there are many Bayesian techniques that can be considered.
Many statisticians consider these techniques to be superior when
applicable.  As we noted, for the applications we have in mind, it is
desirable to assume neither independence nor stationarity, and to
return worst-case quantifications of rejections robust even in
adversarial contexts.  This precludes the application of the standard
approaches.

Nevertheless, when the context is not adversarial and one has high
confidence in the stability of the experimental apparatus and in the
independence of the trials, one can consider applying standard
approaches. But caution is advised when the chosen approach relies on
the central limit theorem.  For the purpose of computing $p$-values,
one needs to compute the worst-case tail probabilities of a test
statistic with respect to all null hypotheses. This computation is 
usually unfeasible unless approximations are used or asymptotics are 
invoked. This becomes an issue when one wishes to make claims of very 
small $p$-values.  An instance of this issue was described in~\cite{Zhang2011}, 
and we reiterate it here. Consider an experimental test of a Bell inequality
$\langle I(X)\rangle \leq 0$. 
At each trial $n$ one observes a value $I(x_n)$ for the random variable $I$.  
In a violating experiment that is perfectly stable and where each trial 
is independent, one expects the estimate $\hat I=\sum_{n=1}^N I(x_n)/N$ to 
converge to a mean $\bar I >0$, where $(\hat I-\bar I)$ converges  
in distribution to a normal distribution with some variance $v$.
It is therefore tempting to compute an estimate $\hat v$ of $v$ from
the data and use the error function to compute a putative $p$-value
$\hat p_N$ from the number of standard deviations of violation
$\hat I/\sqrt{\hat v}$. Note that because $\hat I/\sqrt{\hat v}$ 
is of order $\sqrt{N}$, $\hat p_N$ decreases exponentially with $N$. 
But the central limit theorem cannot be applied here. Convergence 
in distribution implies that for a constant $l$, 
$\textrm{Prob}((\hat I-\bar I)/\sqrt{v}\geq l)$ converges to
the standard normal distribution's probability for this event.  But
for the computation of $\hat p_N$, one needs the probabilities of the events
$(\hat I_\text{LR}-\bar I_\text{LR})/\sqrt{v} \geq \hat I/\sqrt{\hat v}$,
where the subscript ``LR'' means that the estimate and mean are according to 
an LR model. The right-hand side scales as $\sqrt{N}$ and therefore goes 
to infinity as an experiment progresses. Thus, convergence in distribution is 
insufficient for estimating these probabilities. Given this difficulty, one can 
consider other approaches to estimating $p$-values that do not require 
distributional assumptions beyond the parametrized models describing the 
context. A common strategy is to use Monte Carlo sampling according to 
these models, which can work well, particularly for simple null hypotheses.  
We note that it is computationally difficult to use Monte Carlo sampling 
to estimate the very small tail probabilities of interest in the applications 
considered here. These difficulties are avoided by using large deviations 
estimates instead. The martingale-based and PBR protocols can be seen as 
providing such estimates in the absence of independence or stability assumptions.
We remark that many of these protocols' statistical properties can be seen 
as arising from their relationship to test martingales (see Ref.~\cite{Zhang2011} 
for a proof that the sequence of PBRs is a test martingale). Test martingales 
can be used to relate Bayes factors and $p$-values as explained in 
Ref.~\cite{shafer:qc2009a}.

Our comments do not imply that standard statistical approaches cannot
be applied to the problems we are addressing with the PBR protocols. 
However, it is an open problem to determine how to adapt them in contexts 
lacking independence or stationarity and requiring device-independent 
certificates of performance.

\bibliography{simplification}

\begin{thebibliography}{10}

\bibitem{Bell}
J.~S. Bell.
\newblock On the {E}instein {P}odolsky {R}osen paradox.
\newblock {\em Physics}, 1:195--200, 1964.

\bibitem{Genovese}
M.~Genovese.
\newblock Research on hidden variable theories: A review of recent progresses.
\newblock {\em Phys. Rep.}, 413:319--396, 2005.

\bibitem{Barrett2005}
Jonathan Barrett, Lucien Hardy, and Adrian Kent.
\newblock No signaling and quantum key distribution.
\newblock {\em Phys. Rev. Lett.}, 95:010503, Jun 2005.

\bibitem{Masanes2009}
L.~Masanes.
\newblock Universally composable privacy amplification from causality
  constraints.
\newblock {\em Phys. Rev. Lett.}, 102:140501, 2009.

\bibitem{Masanes2011}
Lluis Masanes, Stefano Pironio, and Antonio Acin.
\newblock Secure device-independent quantum key distribution with causally
  independent measurement devices.
\newblock {\em Nat. Commun.}, 2:238, 2011.

\bibitem{Pironio}
S.~Pironio\emph{ et al.}
\newblock Random numbers certified by {B}ell's theorem.
\newblock {\em Nature}, 464:1021, 2010.

\bibitem{Colbeck}
Roger Colbeck and Adrian Kent.
\newblock Private randomness expansion with untrusted devices.
\newblock {\em J. Phys. A: Math. Theor.}, 44:095305, 2011.

\bibitem{Zhang2011}
Yanbao Zhang, Scott Glancy, and Emanuel Knill.
\newblock Asymptotically optimal data analysis for rejecting local realism.
\newblock {\em Phys. Rev. A}, 84:062118, Dec 2011.

\bibitem{Barrett}
Jonathan Barrett, Daniel Collins, Lucien Hardy, Adrian Kent, and Sandu Popescu.
\newblock Quantum nonlocality, {B}ell inequalities, and the memory loophole.
\newblock {\em Phys. Rev. A}, 66(4):042111, Oct 2002.

\bibitem{Gill1}
Richard~D. Gill.
\newblock Accardi contra {B}ell (cum mundi): The impossible coupling.
\newblock In {\em Mathematical Statistics and Applications: Festschrift for
  Constance van Eeden. Eds: M. Moore, S. Froda and C. L\'eger. IMS Lecture
  Notes -- Monograph Series}, volume~42, pages 133--154. Institute of
  Mathematical Statistics. Beachwood, Ohio, 2003.
\newblock Also available as arXiv:quant-ph/0110137.

\bibitem{Gill2}
Richard~D. Gill.
\newblock Time, finite statistics, and {B}ell's fifth position.
\newblock In {\em Proc. of ``Foundations of Probability and Physics - 2", Ser.
  Math. Modelling in Phys., Engin., and Cogn. Sc.}, volume~5, pages 179--206.
  V\"axj\"o Univ. Press., 2003.
\newblock Also available as arXiv:quant-ph/0301059.

\bibitem{Horodecki}
Michal Horodecki, Pawel Horodecki, and Ryszard Horodecki.
\newblock Separability of mixed states: Necessary and sufficient conditions.
\newblock {\em Phys. Lett. A}, 223:1, 1996.

\bibitem{Terhal}
Barbara~M. Terhal.
\newblock Bell inequalities and the separability criterion.
\newblock {\em Phys. Lett. A}, 271:319, 2000.

\bibitem{Clauser}
J.~F. Clauser, M.~A. Horne, A.~Shimony, and R.~A. Holt.
\newblock Proposed experiment to test local hidden-variable theories.
\newblock {\em Phys. Rev. Lett.}, 23:880--884, 1969.

\bibitem{Kullback}
S.~Kullback and R.~A. Leibler.
\newblock On information and sufficiency.
\newblock {\em Ann. Math. Statist.}, 22:79, 1951.

\bibitem{Cover}
T.~M. Cover.
\newblock An algorithm for maximizing expected log investment return.
\newblock {\em IEEE Trans. Inform. Theory}, 30:369, 1984.

\bibitem{Hoeffding}
W.~Hoeffding.
\newblock Probability inequalities for sums of bounded random variables.
\newblock {\em Journ. Amer. Statist. Assoc.}, 58:13, 1963.

\bibitem{mcdiarmid:qc1989a}
C.~McDiarmid.
\newblock On the method of bounded differences.
\newblock In {\em Surveys in Combinatorics}, volume 141 of {\em London Math.
  Soc. Lecture Notes}, pages 148--188. Cambridge Univ. Press, Cambridge, 1989.

\bibitem{Chen}
Jing-Ling Chen, Chunfeng Wu, L.~C. Kwek, C.~H. Oh, and Mo-Lin Ge.
\newblock Violating {B}ell inequalities maximally for two $d$-dimensional
  systems.
\newblock {\em Phys. Rev. A}, 74:032106, Sep 2006.

\bibitem{Collins}
Daniel Collins, Nicolas Gisin, Noah Linden, Serge Massar, and Sandu Popescu.
\newblock Bell inequalities for arbitrarily high-dimensional systems.
\newblock {\em Phys. Rev. Lett.}, 88:040404, Jan 2002.

\bibitem{Bahadur}
R.~R. Bahadur.
\newblock An optimal property of the likelihood ratio statistic.
\newblock In {\em Proc. Fifth Berkeley Symp. on Math. Statist. and Prob.},
  volume~1, pages 13--26. Univ. of Calif. Press, Berkeley, 1967.

\bibitem{Brunner}
Nicolas Brunner, Stefano Pironio, Antonio Acin, Nicolas Gisin, Andr\'e~Allan
  M\'ethot, and Valerio Scarani.
\newblock Testing the dimension of {H}ilbert spaces.
\newblock {\em Phys. Rev. Lett.}, 100:210503, May 2008.

\bibitem{Gallego}
Rodrigo Gallego, Nicolas Brunner, Christopher Hadley, and Antonio Ac\'in.
\newblock Device-independent tests of classical and quantum dimensions.
\newblock {\em Phys. Rev. Lett.}, 105:230501, Nov 2010.

\bibitem{Azuma}
K.~Azuma.
\newblock Weighted sums of certain dependent random variables.
\newblock {\em TohoKu Math. Journ.}, 19:357, 1967.

\bibitem{Robin}
Robin Blume-Kohout, Jun O.~S. Yin, and S.~J. van Enk.
\newblock Entanglement verification with finite data.
\newblock {\em Phys. Rev. Lett.}, 105:170501, Oct 2010.

\bibitem{shafer:qc2009a}
Glenn Shafer, Alexander Shen, Nikolai Vereshchagin, and Vladimir Vovk.
\newblock Test martingales, {B}ayes factors and $p$-values.
\newblock {\em Statist. Sci.}, 26:84--101, 2011.

\end{thebibliography}

\end{document}